\documentclass[11pt,letterpaper]{article}
\usepackage[top=0.85in,left=1in,footskip=0.75in,right=1in]{geometry}

\usepackage{changepage}

\usepackage[utf8]{inputenc}

\usepackage{textcomp,marvosym}

\usepackage{fixltx2e}

\usepackage{amsmath,amssymb}

\usepackage{cite}

\usepackage[breaklinks=true]{hyperref}
\usepackage{nameref}
\usepackage[right]{lineno}

\usepackage{microtype}
\DisableLigatures[f]{encoding = *, family = * }

\usepackage{rotating}

\usepackage[usenames,dvipsnames]{color}

\raggedright
\usepackage[aboveskip=1pt,labelfont=bf,labelsep=period,justification=raggedright,singlelinecheck=off]{caption}

\makeatletter
\renewcommand{\@biblabel}[1]{\quad#1.}
\makeatother

\date{}

\usepackage{lastpage,fancyhdr,graphicx}
\usepackage{epstopdf}
\fancyheadoffset[L]{2.25in}
\fancyfootoffset[L]{2.25in}
\begin{document}
\vspace*{0.35in}

\begin{flushleft}
{\Large
\textbf\newline{Reproducibility of Retinal Thickness Measurements across Spectral-Domain Optical Coherence Tomography Devices using Iowa Reference Algorithm}
}
\newline
\\


Adnan Rashid\textsuperscript{1},
Sebastian M.\ Waldstein\textsuperscript{6},
Bianca S.\ Gerendas\textsuperscript{6},
Hrvoje Bogunovic\textsuperscript{1},
Andreas Wahle\textsuperscript{1},
Kyungmoo Lee\textsuperscript{1},
Kai Wang\textsuperscript{5},
Christian Simader\textsuperscript{6}
Michael D.\ Abramoff\textsuperscript{1,2,3,4}
Ursula Schmidt-Erfurth\textsuperscript{6}
Milan Sonka\textsuperscript{1,2,3}
\\
\bigskip
{1} Electrical and Computer Engineering, The University of Iowa, Iowa City, IA, USA
\\
{2} Ophthalmology and Visual Sciences, The University of Iowa, Iowa City, IA, USA
\\
{3} Biomedical Engineering, The University of Iowa, Iowa City, IA, USA
\\
{4} Veterans Affairs, Medical Center, Iowa City, IA, USA
\\
{5} Department of Biostatistics, The University of Iowa, Iowa City, IA, USA
\\
{6} Christian Doppler Laboratory for Ophthalmic Image Analysis, Vienna Reading Center, Department of Ophthalmology, Medical University of Vienna, Austria
\\
\bigskip

%
%





Corresponding author: milan-sonka@uiowa.edu

\end{flushleft}
\section*{Abstract}
~\indent
PURPOSE: Establishing and obtaining consistent quantitative indices of retinal thickness from a variety of clinically used Spectral-Domain Optical Coherence Tomography scanners.

DESIGN: Retinal images from five Spectral-Domain Optical Coherence Tomography scanners were used to determine total retinal thickness with scanner-specific correction factors establishing consistency of thickness measurement across devices.

PARTICIPANTS: 55 Fovea-centered Spectral-Domain Optical Coherence Tomography volumes from eleven subjects were analyzed, obtained from Cirrus HD-OCT, RS-3000, Heidelberg Spectralis, RTVue and Topcon2000, seven subjects with retinal diseases and four normal controls.

METHOD: The Iowa Reference Algorithm measured total retinal thickness. Nonlinear model of total retinal thickness measurement comparisons was derived and used for device-specific comparisons. Bland-Altman plots and pairwise predictive equations yielded pairwise scanner-specific differences of total retinal thickness. Mendel test determined whether measurement biases were constant for each scanner pair across the cohort of subjects and diseases.

RESULTS: Mendel test revealed that all pairwise scanner differences of total retinal thickness were constant across the cohort (p=0.992). Therefore, individual measurements can be bias-corrected and the Iowa Reference Algorithm serve as a scanner-agnostic independent standard of total retinal thickness across the five tested Spectral-Domain Optical Coherence Tomography scanners.

CONCLUSIONS: Combination of the Iowa Reference Algorithm with scanner-specific bias correction yields cross-scanner consistency of total retinal thickness measurements, facilitating scanning-device independent quantitative assessment of total retinal thickness, longitudinal follow-up quantification without requiring patients to be imaged on the same scanner model, and allowing for multi-center studies with heterogeneous device utilization when using the Iowa Reference Algorithm.



\section*{Introduction}
Since its introduction in 2005, Spectral-Domain Optical Coherence Tomography (SD-OCT)\cite{wojtkowski2005three} has gained rapid acceptance as the imaging tool of choice for the diagnosis, management and treatment of many retinal and neuropathic diseases\cite{abramoff2010retinal,hee1995quantitative,huang1991optical1,puliafito1995imaging}. Measurement of total retinal thickness, between inner limiting membrane (ILM) and Bruch's membrane (BM), is widely used for quantification of disease progression/regression, as well as for diagnostic purposes  when comparison to normative atlases or databases is available. Quantitative measurements of SD-OCT-derived retinal thickness (or thickness of individual retinal layers) are widely used as outcome parameters of large clinical trials such as the Comparison of Age-related macular degeneration Treatments Trial (CATT)\cite{yeh2012uveitis}.

For clinical management of patients imaged in different clinics with different OCT devices, as well as for comparisons in clinical studies, it is important that the quantitative indices obtained from the same eye imaged using different devices at the same time give the same retinal or layer thickness values. However, many studies have reported discrepancies of retinal thickness measurements across devices\cite{wolf2009macular,leung2008comparison,giani2010reproducibility,forooghian2008evaluation,buchser2012comparison,ho2009assessment,huang2009macular}. Such studies, which necessarily use both manufacturer-specific imaging hardware as well as manufacturer/device-specific retinal segmentation algorithms, by nature cannot differentiate whether the observed discrepancies are caused by the imaging hardware (considering scanner resolution, sensor sensitivity, optics, calibration, etc.) or should be attributed to the differences in the segmentation algorithms.

The Iowa Reference Algorithms  are a group of OCT analysis algorithms that have been shown highly reproducible in previously published validation studies\cite{garvin2008intraretinal,lee2010segmentation,li2006optimal,quellec2010three,garvin2009automated,lee20093,bogunovic2014mosaicing}, are publicly available at
\url{<http://www.biomed-imaging.uiowa.edu/downloads>}. They can segment the volumes from essentially any commercial or non-commercial SD-OCT device, including all SD-OCTs that are in routine clinical use. Its flexibility and broad applicability make it an ideal tool for a cross-device comparison study, making the assessment independent of manufacturers implementations of segmentation algorithms. {The purpose of this study is to determine reproducibility of the Iowa Reference Algorithms across five SD-OCT devices, without performing a comparative analysis of segmentation algorithms commercially available on individual OCT scanners.}

In this study, the consistency of retinal layer thickness measurements in normal eyes and eyes with degenerative retinal diseases is assessed using five commonly and widely available SD-OCT devices by  employing the Iowa Reference Algorithm with default parameter setting for total retinal thickness measurement using images from all compared scanners.
\section*{Materials and Methods}
OCT imaging was performed by experienced, reading-center-certified operators using commercially available 3D-OCT 2000 Mark II (Topcon, Tokyo, Japan), Cirrus\textsuperscript{TM} HD-OCT (Carl Zeiss Meditec, Dublin, CA), RS-3000 (Nidek, Tokyo, Japan), RTVue (Optovue, Fremont, CA), and Spectralis\textsuperscript{\textregistered} HRA+OCT (Heidelberg Engineering, Heidelberg, Germany) devices under standardized conditions. Identical or as-similar-as-possible raster scan patterns were selected across all five devices to enable objective comparisons. The study was conducted according to the tenets of the Declaration of Helsinki, and the study protocol was approved by the Institutional Review Board for Human Subjects Research by the ethics committee at the Medical University of Vienna.

All subjects signed informed consent before enrollment. All scans were de-identified before further analysis. The individual scan protocols were as follows:
\begin{itemize}
\item 3D-OCT 2000 (Topcon2000) --- ``3D Macula'' pattern with 128 sections (512 A-scans each) in a 6$\times$6mm area.
\item Cirrus HD-OCT --- ``Macular Cube'' pattern with 128 sections (512 A-scans each) in a 6$\times$6~mm$^2$ area.
\item RS-3000 --- ``Macula Map'' pattern with 128 sections (512 A-scans each) in a 6$\times$6~mm$^2$ area.
\item  RTVue --- ``3D Macular'' pattern with 101 sections (512 A-scans each).
\item Spectralis  --- Custom raster scan pattern with 49 or 25 sections (512 A-scans each; lower number of sections chosen in case of patient fatigue) in a 20$^{\circ}$$\times$20$^{\circ}$ field of view;  automated real-time averaging activated at 29 frames.
\end{itemize}
In all devices, an internal fixation target was provided. In all devices, an internal fixation target was employed. All five scans per subject were completed within one hour in a random acquisition order of OCT devices, to counteract a potential systematic bias and were performed by the same operator.

Table \ref{tab:techdevices} summarizes device-specific image acquisition parameters for all five utilized SD-OCT devices.  RTVue, RS-3000, and Cirrus have an axial resolution of 5$\mu$m.  Axial resolution of Spectralis and Topcon2000 was 4--6$\mu$m and 5--6$\mu$m, respectively. All OCT scans covered the same area of the macula, sized close to 6$\times$6$\times$2~mm$^3$ scan, despite device-specific image resolution differences.

\begin{table}[!ht]
\small
\caption{\bf Image acquisition parameters for the utilized SD-OCT devices.}\label{tab:techdevices}
\centering
\begin{tabular}{|c|c|c|c|c|c|}
\hline
\textbf{SD-OCT} & \textbf{Axial} &  {\textbf{Wavelength}}& \textbf{OCT size} & \textbf{OCT size} & \textbf{Voxel size} \\
\textbf{device} & \textbf{resolution} & && & \\
                & \textbf{($\mu$m)} & {\textbf{(nm)}}&\textbf{(voxels)} & \textbf{(mm$^3$)} & \textbf{($\mu$m$^3$)} \\ \hline
Spectralis      & 4-6  &{870}                         & 512$\times$49/25$\times$496                & 6.00$\times$6.00$\times$1.92         & 11.37$\times$121.20$\times$3.87        \\ \hline
Cirrus        & 5  &{840}                            &512$\times$128$\times$1024	           &6.00$\times$6.00$\times$2.01	        &11.74$\times$46.88$\times$1.96          \\ \hline
RTVue                  & 5 &{840$\pm$10}                             &512$\times$101$\times$640	&6.00$\times$6.00$\times$1.97	 &11.70$\times$59.41$\times$3.08\\ \hline
RS-3000                & 5 &{880}                            &512$\times$128$\times$512	&6.00$\times$6.00$\times$2.09	 &11.70$\times$46.88$\times$4.10\\ \hline
Topcon2000    & 5-6   &{840}                        &512$\times$128$\times$885	&6.00$\times$6.00$\times$2.30	 &11.72$\times$46.88$\times$2.60\\ \hline
\end{tabular}
\end{table}

\begin{figure}[!ht]
        \centering
				\includegraphics[width=0.9\textwidth]{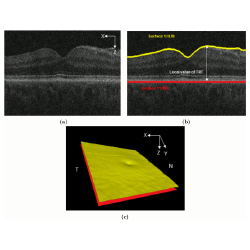}
       \caption{{\bf Total mean retinal thickness is determined as the average of A-scan specific distances between the computer-identified inner limiting membrane and Bruch's membrane (Cirrus macular SD-OCT of a right eye is shown).}			 
(a) Original foveal B-scan image.
(b) Computer segmentation result, inner limiting membrane in yellow,  Bruch's membrane in red.
(c) 3-D rendering of the segmented surfaces  (T: temporal, N: nasal).
}
        \label{SIF:Figure1}
\end{figure}

The Iowa Reference Algorithm\cite{quellec2010three} is a fast multiscale 10-intra-retinal layer (11-surface) segmentation method extended from prior reported work on 3D graph search-based intra-retinal surface segmentation\cite{garvin2009automated,lee20093}, similar approaches based on optimization of graph theory are prevalent in this area \cite{ehnes2014optical}. Briefly, the principal approach is to detect the retinal surfaces in a 3D OCT sub-volume constrained by the retinal surface segmented in a low-resolution 3D OCT volume. The automatically segmented retinal surfaces correspond to cost function minima identified by $s-t$ cut graph search optimization. The cost functions were designed to reflect local dark-to-bright or bright-to-dark image intensity  transitions (Fig. \ref{SIF:Figure1}).
{The Iowa Reference Algorithms are applied to segment all complete 3D SD-OCT scans captured with the 5 devices}. For the purposes of inter-machine comparisons, mean total retinal layer thickness is defined as the average distance along A-scans between Surface 1 (inner limiting membrane (ILM)) and surface 11 (Bruch's membrane (BM)) in the SD-OCT scans, Fig.~\ref{SIF:Figure1}.
In order to calculate the total retinal thickness {for the complete 3D-OCT scans}, all the A-scans ILM--BM distances were represented in micrometers ($\mu$m) by using manufacturer-provided voxel sizes (Table \ref{tab:techdevices}) and averaged.

\subsection*{Statistical analysis}
This paper focuses on reproducibility of assessing total retinal thickness using the above-mentioned five SD-OCT devices. Let $X_{mi}$  denote the mean total retinal thickness measurement from SD-OCT device $m=1,2,...,5$ for subject $i=1,2,...,11$. The segmentation results were reviewed by a retinal specialist (MDA); all ILM and BM surface segmentations were found to be correct. Consequently, the average of all per-subject total retinal thickness measurements obtained from all  SD-OCT devices served as the {internal reference standard}
\begin{equation}\label{eq1}
    X_{ref}(i)=\frac{1}{M}\sum_{m=1}^{M}  X_{mi} \;
\end{equation}
where $M=5$ denotes the number of employed scanners.\\
For each device, a scanner-specific bias was calculated as the mean signed difference between the patient-specific {internal reference standard} and the patient/scanner-specific total retinal thickness measurement
\begin{equation}\label{eq2}
    \overline{X}(m)=\frac{1}{I}\sum_{i=1}^{I}(X_{mi}-X_{ref}(i)) \; ,
\end{equation}
where $I=11$ is the total number of subjects.\\
All reported measurements are expressed in $\mu$m. The scanner-specific bias and the subject-specific bias are reported as mean $\pm$ standard deviation (SD).

Following\cite{carstensen2010comparing}, the model proposed for the comparison of machine-specific total retinal thickness (TRT) measurements $X_{m_i}$ across five SD-OCT devices was
\begin{equation}\label{eq3}
    X_{mi}=\alpha_{m}+\beta_{m}\mu_{i}+\varepsilon_{mi} \; ,
\end{equation}
where $\varepsilon_{mi}$ denotes normal distribution with mean = 0 and variance = 1. This model assumes that the measurement from device $m$ is a linear function of the true value $\mu_i$ with intercept $\alpha_m$ and slope $\beta_m$. The model is more general than classic {\it limits of agreement} analysis in that $\beta_m$ are not required to be the same and  equal to 1. The agreement of measurements from different devices and presence of systematic bias is assessed by Bland-Altman analysis\cite{martin1986statistical}. Using the above model, an approach called Method comparison\cite{carstensen2010comparing} was employed to obtain the prediction equations for each device from a measurement of every other device. The prediction equations were generated using the R package containing MethComp, publicly available  at \url{<http://BendixCarstensen.com/MethComp>}.

Our approach investigates whether the difference in measurements between any two devices is constant. That is, whether the 5 values of $\beta_m$ are the same. The Mendel test\cite{mandel1961non} as implemented in the R package containing additivityTests, publicly available at \url{<https://github.com/rakosnicek/additivityTests>} was used for this purpose.
\begin{figure}[!ht]
        \centering
								\includegraphics[width=0.9\textwidth]{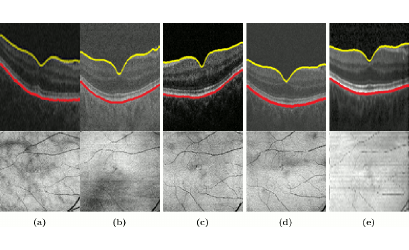}
        \caption{{\bf Examples of middle slices from macular SD-OCT volumes of a normal subject acquired with the five compared  devices.} Retinal layer segmentations (upper panels) and mean total retinal thickness measurements were obtained using the Iowa Reference Algorithm. Lower panels show en-face projection images derived from the respective SD-OCT volumes.
					(a)~Cirrus.
					(b)~RTVue.
					(c)~RS-3000.
					(d)~Topcon2000.
					(e)~Spectralis. }
        \label{SIF:Figure2}
\end{figure}

\section*{Results}
For the purpose of this analysis eleven subjects (eight females, three males) 31 to 80 years old (mean\textsubscript{age} = 58.7, SD\textsubscript{age} = 17.38) were prospectively recruited at the Department of Ophthalmology, Medical University of Vienna, Austria. The study group included four~normal subjects, seven~patients with degenerative retinal diseases (one patient with atrophic age-related macular degeneration (AMD), one macular hole, two Stargardt's disease, three AMD complicated by choroidal neovascularization (CNV)). Fig.~\ref{SIF:Figure2} show example ILM and BM retinal surface segmentations of the middle slices of a normal subject imaged using five SD-OCT devices.
{As already mentioned above, our expert observer (MDA, retinal specialist) considered all segmentations correct with no segmentation failures identified in any of the reviewed B-mode images.}

Bland-Altman plots are shown in Fig.~\ref{SIF:Figure4}, setting the limits of agreement (LOA) to 1.96 times the standard deviations. Table \ref{tab:Tab4} summarizes uncorrected device-specific mean TRT differences, as well as their upper and lower limits  for the five SD-OCT devices.

No device-specific pairwise differences of uncorrected TRT measurements were found statistically different from each other for all five tested SD-OCT devices (Mendel test $p$-value of 0.992). In other words, applying the derived TRT correction model to measurements obtained from individual machine images was expected to yield statistically identical measurements after the correction.

Table \ref{tab:Tab2} summarizes the obtained  mean total retinal thickness measurements for all five SD-OCT devices. The maximum TRT average values were seen for Spectralis (TRT = 292.98 $\pm$ 25.65), while the minimum values per device were observed for RS-3000 (TRT = 276.65 $\pm$ 26.75). RTVue, Cirrus and Topcon2000 TRT measurements ranged between those of Spectralis and RS-3000. Table \ref{tab:Tab2} shows that the inter-subject variability is statistically identical across the five SD-OCT devices.
\begin{table}[!ht]
\caption{\bf Averages of mean retinal thicknesses measured by the Iowa Reference Algorithm across all subject per SD-OCT device.}\label{tab:Tab2}
\small
\centering
\begin{tabular}{l|c|c|c|c|c|}
\cline{2-6}
& \multicolumn{5}{|c|}{\textbf{Mean Total Retinal Thickness  ($\mu$m)}}\\
\cline{2-6}
                            & \multicolumn{1}{|c}{\textbf{Cirrus}} & \multicolumn{1}{|c}{\textbf{RTVue}} & \multicolumn{1}{|c}{\textbf{RS-3000}} & \multicolumn{1}{|c}{\textbf{Spectralis}} & \multicolumn{1}{|c|}{\textbf{Topcon2000}}
\\ \hline
 \multicolumn{1}{|l|}{\textbf{Mean} } &  286.12	&281.92	&276.65	&292.98	&286.30   \\ \cline{1-6}
\multicolumn{1}{|l|}{\textbf{$\pm$SD}}   &26.89	&25.91	&26.75	&25.65	&27.06      \\ \hline
\end{tabular}
\end{table}

\begin{figure}[!ht]
          \centering
									\includegraphics[width=0.9\textwidth]{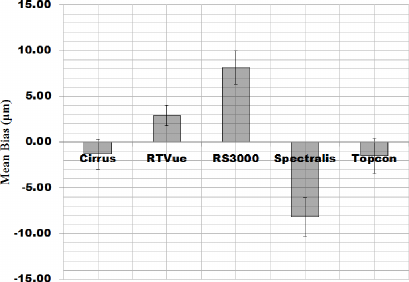}
  \caption{{\bf Device-specific TRT biases determined by comparison with the {internal reference standard}, expressed as mean with $\pm$ 95$\%$ CI (error bars).}}\label{SIF:MeanBias}
\end{figure}

\begin{figure}[!ht]
        \centering
								\includegraphics[width=0.9\textwidth]{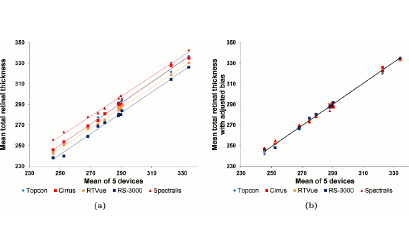}
        \caption{{\bf Plot of device-specific mean total retinal thicknesses of 11 subjects from 5 devices versus the {internal reference standard}; prior to device-specific  bias adjustment (left) and after the bias adjustment (right).}}
        \label{SIF:Figure3}
\end{figure}

\begin{figure}[!ht]
  \begin{center}
					\includegraphics[width=0.9\textwidth]{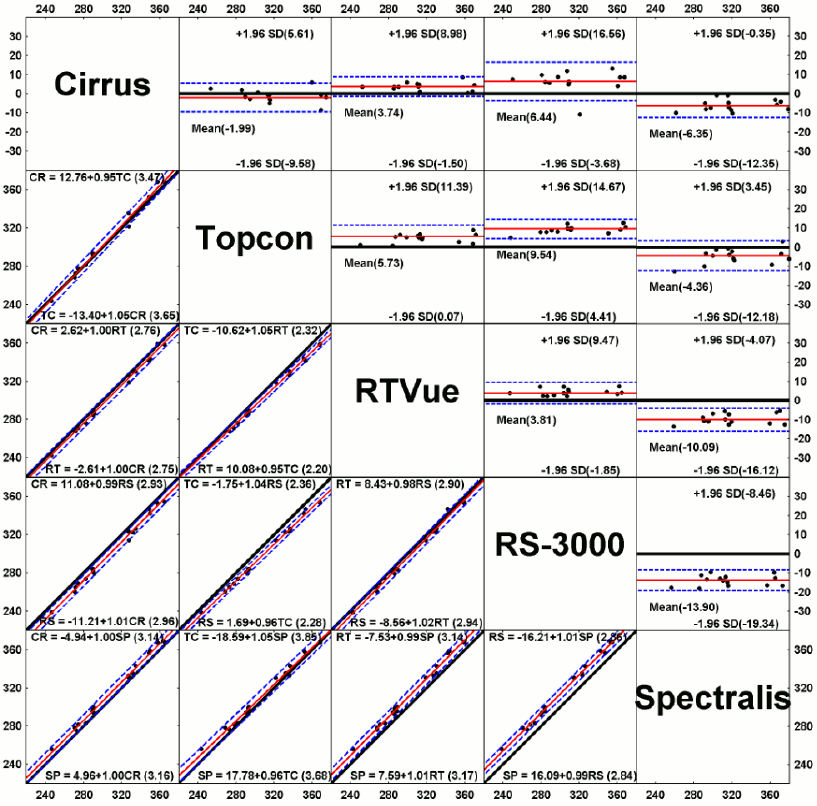}
  \caption{{\bf Pairwise comparisons of TRT measurements from image data acquired by five SD-OCT devices. TRT measurements performed by the Iowa Reference Algorithm. Bland-Altman plots are shown in the upper triangle.} Scatter plots are given in the lower triangle together with the prediction line (red line representing the provided prediction equation), identity line (black line $y=x$) and 95\% upper and lower limits (dotted blue lines).}
  \label{SIF:Figure4}
   \end{center}
\end{figure}

The device-specific bias (expressed as mean\textsubscript{bias}$\pm$SD $\mu$m) is summarized in Table \ref{tab:Tab3}. For total retinal thickness analysis, higher device-specific biases were observed for Spectralis and RS-3000, $-8.19\pm2.14$ $\mu$m and $8.14\pm1.81$ $\mu$m respectively, while RTVue ($2.88\pm1.11$$\mu$m), Cirrus ($-1.33\pm1.64$$\mu$m) and Topcon2000 ($-1.5\pm1.92$$\mu$m) exhibited significantly lower biases. In order to adjust the bias for the subjects, we subtracted the device-specific mean\textsubscript{bias} from the $X_{mi}$ per device. As expected, the corrected TRT measurements per device were statistically similar with low variability, suggesting excellent consistency  of the Iowa Reference Algorithm-derived TRT measurements across the five SD-OCT devices (see Fig.~\ref{SIF:MeanBias}).
In Fig.~\ref{SIF:Figure3}, mean uncorrected and bias-corrected retinal thicknesses $X$ are shown across all five SD-OCT devices for  all subjects in comparison with the {internal reference standard} $X_{ref}$.
\begin{table}[!ht]
\caption{\bf Device-specific TRT biases determined by comparison with the {internal reference standard}, expressed as mean $\pm$ SD.}\label{tab:Tab3}
\small
\centering
\begin{tabular}{l|c|c|c|c|c|}
\cline{2-6}
& \multicolumn{5}{|c|}{\textbf{Mean Total Retinal Thickness  ($\mu$m)}}\\
\cline{2-6}
                            & \multicolumn{1}{|c}{\textbf{Cirrus}} & \multicolumn{1}{|c}{\textbf{RTVue}} & \multicolumn{1}{|c}{\textbf{RS-3000}} & \multicolumn{1}{|c}{\textbf{Spectralis}} & \multicolumn{1}{|c|}{\textbf{Topcon2000}}
\\ \hline
 \multicolumn{1}{|l|}{\textbf{Mean\textsubscript{bias}} } &  -1.33	&2.88	&8.14	&-8.19	 &-1.5   \\ \cline{1-6}
\multicolumn{1}{|l|}{\textbf{$\pm$SD\textsubscript{bias}}}   &1.64	&1.11	&1.81	 &2.14	&1.92     \\ \hline
\end{tabular}
\end{table}

\begin{table}[!ht]
\caption{\bf Mean uncorrected TRT measurements differences and their upper and lower limits for the five SD-OCT devices in $\mu$m.
}\label{tab:Tab4}
\small
 \centering
\begin{tabular}{llccc}
\hline
\textbf{To}                      & \textbf{From}                    & \multicolumn{3}{c}{\textbf{Total Retinal Thickness}}                       \\ \hline
\textbf{Device} & \textbf{Device} & \textbf{MD$\pm$SD} & \textbf{UL:MD+1.96(SD)}       & \textbf{LL:MD-1.96(SD)}
 \\ \hline
Cirrus                           & RTVue                            &04.20$\pm$2.11	    &8.33	 &00.07     \\
Cirrus                           & RS-3000                          &09.47$\pm$2.80	    &14.96	 &03.97        \\
Cirrus                           & Spectralis                       &-06.86$\pm$2.76	&-1.44	 &-12.28 \\
Cirrus                           & Topcon2000                           &-00.18$\pm$3.00	&5.71	 &-06.06   \\
RTVue                            & RS-3000                          &05.27$\pm$2.53	    &10.22	 &00.31        \\
RTVue                            & Spectralis                       &-11.06$\pm$2.43	&-6.30	 &-15.83    \\
RTVue                            & Topcon2000                           &-04.38$\pm$2.21	&-0.05	 &-08.72      \\
RS-3000                          & Spectralis                       &-16.33$\pm$3.34	&-9.79	 &-22.87       \\
RS-3000                          & Topcon2000                           &-09.65$\pm$2.53	&-4.68	 &-14.61        \\
Spectralis                       & Topcon2000                           &06.68$\pm$3.69	    &13.91	 &-00.54               \\ \hline
\multicolumn{5}{l}{\textbf{MD: Mean Diff., UP : Upper Limit and LL: Lower Limit}}      \\ \hline
\end{tabular}
\end{table}

\begin{table}[!ht]
\caption{\bf Predictive equations resulting from the Method comparison between two devices and standard deviation of the prediction error.
}\label{tab:Tab5}
\small
\centering
\begin{tabular}{llcccc}
\hline
\textbf{}       & \textbf{}       & \multicolumn{4}{c}{\textbf{Total Retinal Thickness}}                                                \\ \hline
\textbf{Device} & \textbf{Device} & \textbf{Predictive Equations} & \textbf{$\pm$SD} & \textbf{Predictive Equations} & \textbf{$\pm$SD} \\ \hline
Cirrus          & RTVue           &$CR=-6.41+1.04\times RT$	&2.01	&$RT=6.18+0.96\times CR$	&1.93\\
Cirrus          & RS-3000         &$CR=8.00+1.01\times RS$	&2.96	&$RS=-7.95+0.99\times CR$	&2.94\\
Cirrus          & Spectralis      &$CR=-21.10+1.05\times SP$	&2.66	&$SP=20.13+0.95\times CR$	&2.54\\
Cirrus          & Topcon2000          &$CR=1.62+0.99\times TC$	&3.15	&$TC=-1.63+1.01\times CR$	&3.17\\
RTVue           & RS-3000         &$RT=13.89+0.97\times RS$	&2.48	&$RS=-14.34+1.03\times RT$	&2.56\\
RTVue           & Spectralis      &$RT=-14.15+1.01\times SP$	&2.56	&$SP=14.00+0.99\times RT$	&2.54\\
RTVue           & Topcon2000          &$RT=7.74+0.96\times TC$	&1.95	&$TC=-8.08+1.04\times RT$	&2.04\\
RS-3000         & Spectralis      &$RS=-28.97+1.04\times SP$	&3.39	&$SP=27.77+0.96\times RS$	&3.25\\
RS-3000         & Topcon2000          &$RS=-6.34+0.99\times TC$	&2.63	&$TC=6.42+1.01\times RS$	&2.67\\
Spectralis      & Topcon2000          & $SP=21.70+0.95\times TC$	&3.49	&$TC=-22.90+1.06\times SP$	&3.69 \\ \hline
\end{tabular}
\end{table}
Method comparison\cite{carstensen2010comparing} gives ten predictive equations and predictive errors expressed as standard deviation in Table \ref{tab:Tab5}. Fig.~\ref{SIF:Figure4} shows the Bland-Altman plots (upper triangles) and scatter plots (lower triangles) with predictive equations estimated to illustrate a pair-wise comparison between devices.
\section*{Discussion}
The presented approach facilitates quantitative research using SD-OCT scanners from different manufacturers. This novel advancement is of special importance in longitudinal and/or large enrollment studies.

In this study, we compared the fovea-centered measurements of the total retinal thickness in normal controls as well as in subjects with degenerative retinal diseases. The SD-OCT scans were obtained using five commercially available SD-OCT devices for each subject. The Iowa Reference Algorithm was used to segment retinal surfaces in these volumes. The Iowa Reference Algorithm is  device independent and provides a highly repeatable and reproducible retinal layer segmentation of SD-OCT 3D volumes. In Fig.~\ref{SIF:Figure2}, the delineation of retinal layers by 2 surfaces is shown to be highly similar and stable across the five devices.


In the literature \cite{wolf2009macular,leung2008comparison,giani2010reproducibility,forooghian2008evaluation,buchser2012comparison,ho2009assessment,huang2009macular}, it is argued that the differences of retinal thickness measurements among different SD-OCT devices are associated with different scan acquisition methods, different methods of retinal layer segmentation including alignment and registration, by differences in  sampling frequency of thickness measurement points, and possibly by different estimates of the optical index properties of the retina. Specifically, Wolf et al.\ compared the reproducibility of central retinal thickness (CRT) measurements from six commercially available OCT devices\cite{wolf2009macular}. {In this study, CRT measurements were obtained by scanner-specific assessment of retinal thickness in  normal eyes using following six OCT devices: Stratus OCT (Carl Zeiss Meditec, Inc. Dublin, CA), SOCT Copernicus (Reichert/Optopol Technology, Inc., Depew, NY), Spectral OCT/SLO (Opko/OTI, Inc., Miami, FL), {RTVue-100}, {Spectralis HRA+OCT}, and {Cirrus HD-OCT}. Consistently, higher CRT measurement values were reported by SD-OCT devices compared to TD-OCT scanner. The differences in CRT were substantial and ranged from 2 to 77 $\mu$m. The authors concluded that the major source of the observed discrepancies in CRT measurements across SD-OCT devices was associated with the manufacturer-specific proprietary algorithms for retinal OCT segmentation.}

In another study conducted on normal subjects, similar discrepancies in  macular thickness measurements were reported\cite{leung2008comparison}. Macular thickness measurements were compared between TD-OCT Stratus (Stratus OCT, Carl Zeiss Meditec, Dublin, CA) and SD-OCT (3D OCT, Topcon, Tokyo, Japan). Total and regional macular thicknesses were measured and compared. The mean foveal thicknesses obtained by Stratus OCT and Topcon were 195.6$\pm$17.2 and 260.0$\pm$12.2 $\mu$m, respectively. The total mean macular thickness was 216.4$\pm$18.0$\mu$m (Stratus) and 263.2$\pm$12.6 $\mu$m (Topcon). Macular thickness values obtained from Topcon SD-OCT scanner were generally thicker than those measured using Stratus OCT. In this study, the spans of 95\% limits of agreement for foveal thickness was 21.3 $\mu$m and for total macular thickness was 33.9 $\mu$m.

Mean retinal thickness comparison was reported across 6 OCT devices (one TD-OCT and five SD-OCT): {Stratus, Cirrus, Spectralis HRA-OCT , RTVue-100}, SD-OCT Copernicus HR (Optopol Technology SA, Zawiercie, Poland), and 3D OCT-1000 (Topcon Corporation, Tokyo, Japan) in \cite{giani2010reproducibility}. For all 6 devices, pairwise 95\% limits of agreement (LOA) were not in a clinically acceptable range. The lowest 95\% LOA was 138.9 $\mu$m (Cirrus vs RTVue) and highest 95\% LOA was 348.7 $\mu$m (Copernicus compared to Spectralis). While the authors concluded that causes of the differences in mean retinal thickness were multifactorial, the employed analysis algorithms for retinal layer segmentation, quality of images, and types of pathologies affecting inner-outer retinal layer boundaries were identified as major sources of variability.

Similar findings were reported in an earlier study conducted in  subjects with diabetic macular edema (DME)\cite{forooghian2008evaluation}. Stratus OCT and Cirrus SD-OCT were used to measure and analyze the macular thickness measurements of nine standard macular subfields and total macular volume. For the central macular thickness, the two devices produced measurements that agreed poorly with each other, it was shown that the 95\% limits of agreement were between -137 and +31 $\mu$m.

This work is proposing a new model of TRT measurement comparison across five SD-OCT devices. In a conventional limit-of-agreement analysis for comparing two measurements methods, the assumption is that parameters $\beta_m$ are identical or equal to 1. In our approach, this assumption is not required yielding an assessment approach that is generally applicable to comparisons of two or more measurement methods. The  Mendel test applied to the proposed model reveals that the five respective parameters $\beta_m$ ($m=1,...,5$) expressing the pairwise differences among the TRT measurements across SD-OCT devices are statistically indistinguishable ($p=0.992$). Having established this, we estimated a signed device-specific bias for each SD-OCT device, investigated the Bland-Altman plots for a proportional error and/or variation of measurements depending strongly on the magnitude of measurements, and also presented ten predictive equations in a pair-wise manner for the conversion of TRT measurements across five SD-OCT devices.

Table \ref{tab:techdevices} details a signed device-specific bias for each SD-OCT device. The maximum device-specific bias of $-8.19\pm2.14$ $\mu$m was observed for Spectralis compared to other devices -- the variability of the measured device-specific biases ranged from $\pm1.11$ to $\pm2.14$ $\mu$m, which is significantly lower than the device-specific voxel sizes (Table \ref{tab:Tab3}, Fig.~\ref{SIF:MeanBias} and Table \ref{tab:techdevices}). Furthermore, we have shown that excellent cross-device reproducibility can be obtained by device-specific bias correction -- mean retinal thickness measurements from different devices are tightly clustered around the same value for each subject once subjected to the correction factor (Fig.~\ref{SIF:Figure3}).

Bland-Altman plots were used to compare five SD-OCT devices in pairwise comparisons (Table \ref{SIF:Figure4}). Generally, the Bland-Altman plots (Fig.~\ref{SIF:Figure4}) show no typical trends.
Despite its unquestionable importance, the presented study is not free of several limitations.
First, the cross-scanner consistency was demonstrated on a relatively small sample of 11 subjects. Yet, the imaging protocol required each subject to be imaged 5 times on the same day using five different OCT scanners, limiting the sample size. Additionally, the enrolled subjects included a mixture of pathologies and normal subjects as described earlier. As a result, the analyzed dataset is based on 55 volumetric OCT images, altogether consisting of 5874 OCT image slices. The image acquisition protocol is therefore unique and resulted in a highly diverse set of OCT image data.

Another limitation of the study design is its reliance on total retinal thickness. As was presented in earlier papers from our group, the Iowa Reference Algorithm routinely generates 11 retinal surfaces and thus outlines 10 inner- and outer-retinal layers. We have decided to limit the cross-machine comparison to total retinal thickness for two reasons: 1)~10-layer OCT segmentation in retinas suffering from topology-disrupting pathologies (e.g., cysts as in DME, AMD, etc.) remains an unsolved problem -- consequently, not all layers may always be segmented reliably in such pathologies and 2)~a wide range of thickness values was necessary to derive a reliable set of device-specific measurement correction parameters.
While a study could have been designed using normal subjects (and/or subjects with diseases not disrupting retinal layer topology) with all 10 layers reliably segmented, the ranges of layer-specific thicknesses would be small and consequently the analysis resulting in robust correction factors would not be possible.

Once a larger dataset of OCT images from different scanners becomes available, one of the future goals may be to specifically focus on developing a correction approach for total retinal thickness. The presented method may then offer machine specific bias correction, hopefully with a high consistency across devices. The presented approach will in that case also lead naturally and directly to correcting thickness measurements of individual retinal layers since the same correction factors would apply.
At this stage, while the presented results appear promising, no claims can be made that the Iowa Reference Algorithm is actually superior to using OCT device implemented algorithms in multicenter studies. Differing results in previous studies might have been the result of other factors like measurement technique etc.
Nevertheless, this study holds a promise for eventually achieving accurate multi-device cross-scanner studies of retinal morphology.

\section*{Conclusion}
The Iowa Reference Algorithm offers a means of obtaining repeatable measurements of retinal thickness across clinically-relevant SD-OCT devices (Cirrus, Spectralis, Topcon2000, RTVue, and RS-3000). {The Bland-Altman analysis of the device-bias-uncorrected TRT measurements demonstrate that once the device-specific correction factors are established on a larger dataset of OCT images from a variety of OCT scanners, interchangeability between the considered devices may be possible when proper corrections are applied.} Predictive equations provided above advocate a strategy for such pairwise device measurement conversions. The observed systemic errors suggest presence of a measurable device-specific bias among all five studied SD-OCT devices.
The measurement variance of around $\approx$ 2 $\mu$m (less than the device-specific axial voxel sizes) in bias estimation and the achieved $p$-value of 0.992 with the proposed model for 5 separate values of $\beta_m$ for SD-OCTs suggest the cross-scanner reproducibility of the Iowa Reference Algorithm post-correction measurements and thus the high reproducibility of its TRT measurements. {However, this pilot study is too limited to establish authoritative correction factors across the different devices. A larger study, which we are currently planning, is required to do that. The presented results hold promise for future multi-center studies with heterogeneous device utilization.}

\section*{Acknowledgments}
This work was partially supported by The Christian Doppler Society OPTIMA project; NIH grants R01 EY019112, R01 EY018853 and R01 EB004640; the Department of Veterans Affairs; the Marlene S.\ and Leonard A.\ Hadley Glaucoma Research Fund; and the
Arnold and Mabel Beckman Initiative for Macular Research (BIMR) \#18350500.

\nolinenumbers

%
%
%
\bibliography{mybibfile}

\begin{thebibliography}{10}

\bibitem{abramoff2010retinal}
Michael~D Abramoff, Mona~K Garvin, and Milan Sonka.
\newblock Retinal imaging and image analysis.
\newblock {\em IEEE Rev Biomed Eng.}, 3:169--208, 2010.

\bibitem{bogunovic2014mosaicing}
H.~Bogunovic, M.~Sonka, Y.~H. Kwon, P.~Kemp, M.~D. Abramoff, and X.~Wu.
\newblock {{M}ulti-surface and multi-field co-segmentation of 3-{D} retinal
  optical coherence tomography}.
\newblock {\em IEEE Trans Med Imaging}, 33(12):2242--2253, 2014.

\bibitem{buchser2012comparison}
Nancy~M Buchser, Gadi Wollstein, Hiroshi Ishikawa, Richard~A Bilonick, Yun
  Ling, Lindsey~S Folio, Larry Kagemann, Robert~J Noecker, Eiyass Albeiruti,
  and Joel~S Schuman.
\newblock Comparison of retinal nerve fiber layer thickness measurement bias
  and imprecision across three spectral-domain optical coherence tomography
  devices.
\newblock {\em Invest Ophthalmol Vis Sci.}, 53:3742--3747, 2012.

\bibitem{carstensen2010comparing}
Bendix Carstensen.
\newblock Comparing methods of measurement: Extending the loa by regression.
\newblock {\em Stat Med.}, 29:401--410, 2010.

\bibitem{ehnes2014optical}
A.~Ehnes, Y.~Wenner, C.~Friedburg, M.~N. Preising, W.~Bowl, W.~Sekundo, E.~M.
  Zu~Bexten, K.~Stieger, and B.~Lorenz.
\newblock {{O}ptical {C}oherence {T}omography ({O}{C}{T}) {D}evice
  {I}ndependent {I}ntraretinal {L}ayer {S}egmentation}.
\newblock {\em Transl Vis Sci Technol}, 3(1):1, 2014.

\bibitem{forooghian2008evaluation}
Farzin Forooghian, Catherine Cukras, Catherine~B Meyerle, Emily~Y Chew, and
  Wai~T Wong.
\newblock Evaluation of time domain and spectral domain optical coherence
  tomography in the measurement of diabetic macular edema.
\newblock {\em Invest Ophthalmol Vis Sci.}, 49:4290--4296, 2008.

\bibitem{garvin2008intraretinal}
Mona~Kathryn Garvin, Michael~D Abr{\`a}moff, Randy Kardon, Stephen~R Russell,
  Xiaodong Wu, and Milan Sonka.
\newblock Intraretinal layer segmentation of macular optical coherence
  tomography images using optimal 3-{D} graph search.
\newblock {\em IEEE Trans Med Imaging.}, 27:1495--1505, 2008.

\bibitem{garvin2009automated}
Mona~Kathryn Garvin, Michael~D Abr{\`a}moff, Xiaodong Wu, Stephen~R Russell,
  Trudy~L Burns, and Milan Sonka.
\newblock Automated 3-{D} intraretinal layer segmentation of macular
  spectral-domain optical coherence tomography images.
\newblock {\em IEEE Trans Med Imaging.}, 28:1436--1447, 2009.

\bibitem{giani2010reproducibility}
Andrea Giani, Mario Cigada, Netan Choudhry, Antonio~Peroglio Deiro, Marta
  Oldani, Marco Pellegrini, Alessandro Invernizzi, Piergiorgio Duca, Joan~W
  Miller, and Giovanni Staurenghi.
\newblock Reproducibility of retinal thickness measurements on normal and
  pathologic eyes by different optical coherence tomography instruments.
\newblock {\em Am J Ophthalmol.}, 150:815--824, 2010.

\bibitem{hee1995quantitative}
Michael~R Hee, Carmen~A Puliafito, Carlton Wong, Jay~S Duker, Elias Reichel,
  Bryan Rutledge, Joel~S Schuman, Eric~A Swanson, and James~G Fujimoto.
\newblock Quantitative assessment of macular edema with optical coherence
  tomography.
\newblock {\em Arch Ophthalmol.}, 113:1019--1029, 1995.

\bibitem{ho2009assessment}
Joseph Ho, Alan~C Sull, Laurel~N Vuong, Yueli Chen, Jonathan Liu, James~G
  Fujimoto, Joel~S Schuman, and Jay~S Duker.
\newblock Assessment of artifacts and reproducibility across spectral-and
  time-domain optical coherence tomography devices.
\newblock {\em Ophthalmology.}, 116:1960--1970, 2009.

\bibitem{huang1991optical1}
David Huang, Eric~A Swanson, Charles~P Lin, Joel~S Schuman, William~G Stinson,
  Warren Chang, Michael~R Hee, Thomas Flotte, Kenton Gregory, Carmen~A
  Puliafito, et~al.
\newblock Optical coherence tomography.
\newblock {\em Science.}, 254:1178--1181, 1991.

\bibitem{huang2009macular}
Jingjing Huang, Xing Liu, Ziqiang Wu, Hui Xiao, Laurie Dustin, and Srinivas
  Sadda.
\newblock Macular thickness measurements in normal eyes with time domain and
  fourier domain optical coherence tomography.
\newblock {\em Retina.}, 29:980, 2009.

\bibitem{lee20093}
Kyungmoo Lee, Meindert Niemeijer, Mona~K Garvin, Young~H Kwon, Milan Sonka, and
  Michael~D Abr{\`a}moff.
\newblock 3-{D} segmentation of the rim and cup in spectral-domain optical
  coherence tomography volumes of the optic nerve head.
\newblock In {\em SPIE Medical Imaging.}, pages 72622D--72622D. International
  Society for Optics and Photonics, 2009.

\bibitem{lee2010segmentation}
Kyungmoo Lee, Meindert Niemeijer, Mona~K Garvin, Young~H Kwon, Milan Sonka, and
  Michael~D Abr{\`a}moff.
\newblock Segmentation of the optic disc in 3-{D} {OCT} scans of the optic
  nerve head.
\newblock {\em IEEE Trans Med Imaging.}, 29:159--168, 2010.

\bibitem{leung2008comparison}
Christopher Kai-shun Leung, Carol Yim-lui Cheung, Robert~N Weinreb, Gary Lee,
  Dusheng Lin, Chi~Pui Pang, and Dennis~SC Lam.
\newblock Comparison of macular thickness measurements between time domain and
  spectral domain optical coherence tomography.
\newblock {\em Invest Ophthalmol Vis Sci.}, 49:4893--4897, 2008.

\bibitem{li2006optimal}
Kang Li, Xiaodong Wu, Danny~Z Chen, and Milan Sonka.
\newblock Optimal surface segmentation in volumetric images-a graph-theoretic
  approach.
\newblock {\em IEEE Trans. Pattern Anal.}, 28:119--134, 2006.

\bibitem{mandel1961non}
John Mandel.
\newblock Non-additivity in two-way analysis of variance.
\newblock {\em J Am Statist Assoc.}, 56:878--888, 1961.

\bibitem{martin1986statistical}
J~Martin~Bland and DouglasG Altman.
\newblock Statistical methods for assessing agreement between two methods of
  clinical measurement.
\newblock {\em Lancet.}, 327:307--310, 1986.

\bibitem{puliafito1995imaging}
Carmen~A Puliafito, Michael~R Hee, Charles~P Lin, Elias Reichel, Joel~S
  Schuman, Jay~S Duker, Joseph~A Izatt, Eric~A Swanson, and James~G Fujimoto.
\newblock Imaging of macular diseases with optical coherence tomography.
\newblock {\em Ophthalmology.}, 102:217--229, 1995.

\bibitem{quellec2010three}
Gw{\'e}nol{\'e} Quellec, Kyungmoo Lee, Martin Dolejsi, Mona~Kathryn Garvin,
  Michael~D Abr{\`a}moff, and Milan Sonka.
\newblock Three-dimensional analysis of retinal layer texture: identification
  of fluid-filled regions in {SD-OCT} of the macula.
\newblock {\em IEEE Trans Med Imaging.}, 29:1321--1330, 2010.

\bibitem{wojtkowski2005three}
Maciej Wojtkowski, Vivek Srinivasan, James~G Fujimoto, Tony Ko, Joel~S Schuman,
  Andrzej Kowalczyk, and Jay~S Duker.
\newblock Three-dimensional retinal imaging with high-speed
  ultrahigh-resolution optical coherence tomography.
\newblock {\em Ophthalmology.}, 112:1734--1746, 2005.

\bibitem{wolf2009macular}
Ute~EK Wolf-Schnurrbusch, Lala Ceklic, Christian~K Brinkmann, Milko~E Iliev,
  Manuel Frey, Simon~P Rothenbuehler, Volker Enzmann, and Sebastian Wolf.
\newblock Macular thickness measurements in healthy eyes using six different
  optical coherence tomography instruments.
\newblock {\em Invest Ophthalmol Vis Sci.}, 50:3432--3437, 2009.

\bibitem{yeh2012uveitis}
Steven Yeh, Thomas~A Albini, Andrew~A Moshfeghi, and Robert~B Nussenblatt.
\newblock Uveitis, the comparison of age-related macular degeneration
  treatments trials (catt), and intravitreal biologics for ocular inflammation.
\newblock {\em Am J Ophthalmol.}, 154:429--435, 2012.

\end{thebibliography}

\end{document}